

\magnification=\magstep1

\newbox\SlashedBox
\def\slashed#1{\setbox\SlashedBox=\hbox{#1}
\hbox to 0pt{\hbox to 1\wd\SlashedBox{\hfil/\hfil}\hss}#1}
\def\hboxtosizeof#1#2{\setbox\SlashedBox=\hbox{#1}
\hbox to 1\wd\SlashedBox{#2}}

\def\mathslashed#1{\setbox\SlashedBox=\hbox{$#1$}
\hbox to 0pt{\hbox to 1\wd\SlashedBox{\hfil/\hfil}\hss}#1}

\def\ifsmall{\iffalse}  
\def\titlepagefont{}  

\def\DefineTeXgraphics{%
\special{ps::[global] /TeXgraphics { } def}}  

\def\today{\ifcase\month\or January\or February\or March\or April\or May
\or June\or July\or August\or September\or October\or November\or
December\fi\space\number\day, \number\year}
\def\eatPrefix19{}
\def\Year{\expandafter\eatPrefix\the\year}
\newcount\hours \newcount\minutes
\def\monthname{\ifcase\month\or
January\or February\or March\or April\or May\or June\or July\or
August\or September\or October\or November\or December\fi}
\def\shortmonthname{\ifcase\month\or
Jan\or Feb\or Mar\or Apr\or May\or Jun\or Jul\or
Aug\or Sep\or Oct\or Nov\or Dec\fi}

\def\TimeStamp{\hours\the\time\divide\hours by60%
\minutes -\the\time\divide\minutes by60\multiply\minutes by60%
\advance\minutes by\the\time%
${\rm \shortmonthname}\cdot\if\day<10{}0\fi\the\day\cdot\the\year%
\qquad\the\hours:\if\minutes<10{}0\fi\the\minutes$}







\newif\ifdraftmode
\newif\ifleftlabels  

\def\nolabels{\def\wrlabeL##1{}\def\eqlabeL##1{}\def\reflabeL##1{}}
\def\writelabels{\def\wrlabeL##1{\leavevmode\vadjust{\rlap{\smash%
{\line{{\escapechar=` \hfill\rlap{\sevenrm\hskip.03in\string##1}}}}}}}%
\def\eqlabeL##1{{\escapechar-1\rlap{\sevenrm\hskip.05in\string##1}}}%
\def\reflabeL##1{\noexpand\rlap{\noexpand\sevenrm[\string##1]}}}
\def\writeleftlabels{\def\wrlabeL##1{\leavevmode\vadjust{\rlap{\smash%
{\line{{\escapechar=` \hfill\rlap{\sevenrm\hskip.03in\string##1}}}}}}}%
\def\eqlabeL##1{{\escapechar-1%
\rlap{\sixrm\hskip.05in\string##1}%
\llap{\sevenrm\string##1\hskip.03in\hbox to \hsize{}}}}%
\def\reflabeL##1{\noexpand\rlap{\noexpand\sevenrm[\string##1]}}}
\nolabels

\newdimen\fullhsize
\newdimen\hstitle
\hstitle=\hsize 
\newdimen\hsbody
\hsbody=\hsize 
\newdimen\hbodyoffset
\hbodyoffset=\hoffset 
\newbox\leftpage
\def\abstract#1{#1}
\def\rotated{\special{ps: landscape}
\magnification=1000  
\baselineskip=14pt
\global\hstitle=9truein\global\hsbody=4.75truein
\global\vsize=7truein\global\voffset=-.31truein
\global\hoffset=-0.54in\global\hbodyoffset=-.54truein
\global\fullhsize=10truein
\def\DefineTeXgraphics{%
\special{ps::[global]
/TeXgraphics {currentpoint translate 0.7 0.7 scale
              -80 0.72 mul -1000 0.72 mul translate} def}}
\let\lr=L
\def\ifsmall{\iftrue}
\def\titlepagefont{\twelvepoint}
\trueseventeenpoint
\def\almostshipout##1{\if L\lr \count1=1
      \global\setbox\leftpage=##1 \global\let\lr=R
   \else \count1=2
      \shipout\vbox{\hbox to\fullhsize{\box\leftpage\hfil##1}}
      \global\let\lr=L\fi}

\output={\ifnum\count0=1 
 \shipout\vbox{\hbox to \fullhsize{\hfill\pagebody\hfill}}\advancepageno
 \else
 \almostshipout{\leftline{\vbox{\pagebody\makefootline}}}\advancepageno
 \fi}

\def\abstract##1{{\leftskip=1.5in\rightskip=1.5in ##1\par}} }

\def\linemessage#1{\immediate\write16{#1}}

\global\newcount\secno \global\secno=0
\global\newcount\appno \global\appno=0
\global\newcount\meqno \global\meqno=1
\global\newcount\subsecno \global\subsecno=0
\global\newcount\figno \global\figno=0

\newif\ifAnyCounterChanged
\let\terminator=\relax
\def\normalize#1{\ifx#1\terminator\let\next=\relax\else%
\if#1i\aftergroup i\else\if#1v\aftergroup v\else\if#1x\aftergroup x%
\else\if#1l\aftergroup l\else\if#1c\aftergroup c\else%
\if#1m\aftergroup m\else%
\if#1I\aftergroup I\else\if#1V\aftergroup V\else\if#1X\aftergroup X%
\else\if#1L\aftergroup L\else\if#1C\aftergroup C\else%
\if#1M\aftergroup M\else\aftergroup#1\fi\fi\fi\fi\fi\fi\fi\fi\fi\fi\fi\fi%
\let\next=\normalize\fi%
\next}
\def\makeNormal#1#2{\def\doNormalDef{\edef#1}\begingroup%
\aftergroup\doNormalDef\aftergroup{\normalize#2\terminator\aftergroup}%
\endgroup}

\def\warnIfChanged#1#2{%
\ifundef#1
\else\begingroup%
\edef\oldDefinitionOfCounter{#1}\edef\newDefinitionOfCounter{#2}%
\ifx\oldDefinitionOfCounter\newDefinitionOfCounter%
\else%
\linemessage{Warning: definition of \noexpand#1 has changed.}%
\global\AnyCounterChangedtrue\fi\endgroup\fi}

\def\Section#1{\global\advance\secno by1\relax\global\meqno=1%
\global\subsecno=0%
\bigbreak\bigskip
\centerline{\twelvepoint \bf %
\the\secno. #1}%
\par\nobreak\medskip\nobreak}
\def\tagsection#1{%
\warnIfChanged#1{\the\secno}%
\xdef#1{\the\secno}%
\ifWritingAuxFile\immediate\write\auxfile{\noexpand\xdef\noexpand#1{#1}}\fi%
}
\def\section{\Section}
\def\Subsection#1{\global\advance\subsecno by1\relax\medskip %
\leftline{\bf\the\secno.\the\subsecno\ #1}%
\par\nobreak\smallskip\nobreak}
\def\tagsubsection#1{%
\warnIfChanged#1{\the\secno.\the\subsecno}%
\xdef#1{\the\secno.\the\subsecno}%
\ifWritingAuxFile\immediate\write\auxfile{\noexpand\xdef\noexpand#1{#1}}\fi%
}

\def\subsection{\Subsection}

\def\romappno{\uppercase\expandafter{\romannumeral\appno}}
\def\makeNormalizedRomappno{%
\expandafter\makeNormal\expandafter\normalizedromappno%
\expandafter{\romannumeral\appno}%
\edef\normalizedromappno{\uppercase{\normalizedromappno}}}
\def\Appendix#1{\global\advance\appno by1\relax\global\meqno=1\global\secno=0%
\global\subsecno=0%
\bigbreak\bigskip
\centerline{\twelvepoint \bf Appendix %
\romappno. #1}%
\par\nobreak\medskip\nobreak}
\def\tagappendix#1{\makeNormalizedRomappno%
\warnIfChanged#1{\normalizedromappno}%
\xdef#1{\normalizedromappno}%
\ifWritingAuxFile\immediate\write\auxfile{\noexpand\xdef\noexpand#1{#1}}\fi%
}
\def\appendix{\Appendix}
\def\Subappendix#1{\global\advance\subsecno by1\relax\medskip %
\leftline{\bf\romappno.\the\subsecno\ #1}%
\par\nobreak\smallskip\nobreak}
\def\tagsubappendix#1{\makeNormalizedRomappno%
\warnIfChanged#1{\normalizedromappno.\the\subsecno}%
\xdef#1{\normalizedromappno.\the\subsecno}%
\ifWritingAuxFile\immediate\write\auxfile{\noexpand\xdef\noexpand#1{#1}}\fi%
}

\def\eqn#1{\makeNormalizedRomappno%
\ifnum\secno>0%
  \warnIfChanged#1{\the\secno.\the\meqno}%
  \eqno(\the\secno.\the\meqno)\xdef#1{\the\secno.\the\meqno}%
     \global\advance\meqno by1
\else\ifnum\appno>0%
  \warnIfChanged#1{\normalizedromappno.\the\meqno}%
  \eqno({\rm\romappno}.\the\meqno)%
      \xdef#1{\normalizedromappno.\the\meqno}%
     \global\advance\meqno by1
\else%
  \warnIfChanged#1{\the\meqno}%
  \eqno(\the\meqno)\xdef#1{\the\meqno}%
     \global\advance\meqno by1
\fi\fi%
\eqlabeL#1%
\ifWritingAuxFile\immediate\write\auxfile{\noexpand\xdef\noexpand#1{#1}}\fi%
}
\def\defeqn#1{\makeNormalizedRomappno%
\ifnum\secno>0%
  \warnIfChanged#1{\the\secno.\the\meqno}%
  \xdef#1{\the\secno.\the\meqno}%
     \global\advance\meqno by1
\else\ifnum\appno>0%
  \warnIfChanged#1{\normalizedromappno.\the\meqno}%
  \xdef#1{\normalizedromappno.\the\meqno}%
     \global\advance\meqno by1
\else%
  \warnIfChanged#1{\the\meqno}%
  \xdef#1{\the\meqno}%
     \global\advance\meqno by1
\fi\fi%
\eqlabeL#1%
\ifWritingAuxFile\immediate\write\auxfile{\noexpand\xdef\noexpand#1{#1}}\fi%
}
\def\anoneqn{\makeNormalizedRomappno%
\ifnum\secno>0
  \eqno(\the\secno.\the\meqno)%
     \global\advance\meqno by1
\else\ifnum\appno>0
  \eqno({\rm\normalizedromappno}.\the\meqno)%
     \global\advance\meqno by1
\else
  \eqno(\the\meqno)%
     \global\advance\meqno by1
\fi\fi%
}
\def\mfig#1#2{\global\advance\figno by1%
\relax#1\the\figno%
\warnIfChanged#2{\the\figno}%
\edef#2{\the\figno}%
\reflabeL#2%
\ifWritingAuxFile\immediate\write\auxfile{\noexpand\xdef\noexpand#2{#2}}\fi%
}

\def\fig#1{\mfig{fig.\ ~}#1}

\catcode`@=11 

\newif\ifFiguresInText\FiguresInTexttrue
\newif\if@FigureFileCreated
\newwrite\capfile
\newwrite\figfile

\def\PlaceTextFigure#1#2#3#4{%
\vskip 0.5truein%
#3\hfil\epsfbox{#4}\hfil\break%
\hfil\vbox{Figure #1. #2}\hfil%
\vskip10pt}
\def\PlaceEndFigure#1#2{%
\epsfysize=\vsize\epsfbox{#2}\hfil\break\vfill\centerline{Figure #1.}\eject}

\def\LoadFigure#1#2#3#4{%
\ifundef#1{\phantom{\mfig{}#1}}\fi
\ifWritingAuxFile\immediate\write\auxfile{\noexpand\xdef\noexpand#1{#1}}\fi%
\ifFiguresInText
\PlaceTextFigure{#1}{#2}{#3}{#4}%
\else
\if@FigureFileCreated\else%
\immediate\openout\capfile=\jobname.caps%
\immediate\openout\figfile=\jobname.figs%
\fi%
\immediate\write\capfile{\noexpand\item{Figure \noexpand#1.\ }#2.}%
\immediate\write\figfile{\noexpand\PlaceEndFigure\noexpand#1{\noexpand#4}}%
\fi}

\def\listfigs{\ifFiguresInText\else%
\vfill\eject\immediate\closeout\capfile
\immediate\closeout\figfile%
\centerline{{\bf Figures}}\bigskip\frenchspacing%
\input \jobname.caps\vfill\eject\nonfrenchspacing%
\input\jobname.figs\fi}

\font\ninerm=cmr9
\font\eightrm=cmr8
\font\sixrm=cmr6

\def\loadtrueseventeenpoint{
 \font\seventeenrm=cmr10 at 17.28truept
 \font\seventeeni=cmmi10 at 17.28truept
 \font\seventeenbf=cmbx10 at 17.28truept
 \font\seventeenit=cmti10 at 17.28truept
 \font\seventeensl=cmsl10 at 17.28truept
 \font\seventeensy=cmsy10 at 17.28truept
}
\def\loadfourteenpoint{
\font\fourteenrm=cmr10 at 14.4pt
\font\fourteeni=cmmi10 at 14.4pt
\font\fourteenit=cmti10 at 14.4pt
\font\fourteensl=cmsl10 at 14.4pt
\font\fourteensy=cmsy10 at 14.4pt
\font\fourteenbf=cmbx10 at 14.4pt
}
\def\loadtruetwelvepoint{
\font\twelverm=cmr10 at 12truept
\font\twelvei=cmmi10 at 12truept
\font\twelveit=cmti10 at 12truept
\font\twelvesl=cmsl10 at 12truept
\font\twelvesy=cmsy10 at 12truept
\font\twelvebf=cmbx10 at 12truept
}

\font\ninei=cmmi9
\font\eighti=cmmi8
\font\sixi=cmmi6
\skewchar\ninei='177 \skewchar\eighti='177 \skewchar\sixi='177

\font\ninesy=cmsy9
\font\eightsy=cmsy8
\font\sixsy=cmsy6
\skewchar\ninesy='60 \skewchar\eightsy='60 \skewchar\sixsy='60

\font\ninebf=cmbx9
\font\eightbf=cmbx8
\font\sixbf=cmbx6

\font\ninett=cmtt9
\font\eighttt=cmtt8

\hyphenchar\tentt=-1 
\hyphenchar\ninett=-1
\hyphenchar\eighttt=-1

\font\ninesl=cmsl9
\font\eightsl=cmsl8

\font\nineit=cmti9
\font\eightit=cmti8


\newskip\ttglue
\def\tenpoint{\def\rm{\fam0\tenrm}%
  \textfont0=\tenrm \scriptfont0=\sevenrm \scriptscriptfont0=\fiverm
  \textfont1=\teni \scriptfont1=\seveni \scriptscriptfont1=\fivei
  \textfont2=\tensy \scriptfont2=\sevensy \scriptscriptfont2=\fivesy
  \textfont3=\tenex \scriptfont3=\tenex \scriptscriptfont3=\tenex
  \def\it{\fam\itfam\tenit}\textfont\itfam=\tenit
  \def\sl{\fam\slfam\tensl}\textfont\slfam=\tensl
  \def\bf{\fam\bffam\tenbf}\textfont\bffam=\tenbf \scriptfont\bffam=\sevenbf
  \scriptscriptfont\bffam=\fivebf
  \normalbaselineskip=12pt
  \let\sc=\eightrm
  \let\big=\tenbig
  \setbox\strutbox=\hbox{\vrule height8.5pt depth3.5pt width\z@}%
  \normalbaselines\rm}

\def\twelvepoint{\def\rm{\fam0\twelverm}%
  \textfont0=\twelverm \scriptfont0=\ninerm \scriptscriptfont0=\sevenrm
  \textfont1=\twelvei \scriptfont1=\ninei \scriptscriptfont1=\seveni
  \textfont2=\twelvesy \scriptfont2=\ninesy \scriptscriptfont2=\sevensy
  \textfont3=\tenex \scriptfont3=\tenex \scriptscriptfont3=\tenex
  \def\it{\fam\itfam\twelveit}\textfont\itfam=\twelveit
  \def\sl{\fam\slfam\twelvesl}\textfont\slfam=\twelvesl
  \def\bf{\fam\bffam\twelvebf}\textfont\bffam=\twelvebf%
  \scriptfont\bffam=\ninebf
  \scriptscriptfont\bffam=\sevenbf
  \normalbaselineskip=12pt
  \let\sc=\eightrm
  \let\big=\tenbig
  \setbox\strutbox=\hbox{\vrule height8.5pt depth3.5pt width\z@}%
  \normalbaselines\rm}

\def\fourteenpoint{\def\rm{\fam0\fourteenrm}%
  \textfont0=\fourteenrm \scriptfont0=\tenrm \scriptscriptfont0=\sevenrm
  \textfont1=\fourteeni \scriptfont1=\teni \scriptscriptfont1=\seveni
  \textfont2=\fourteensy \scriptfont2=\tensy \scriptscriptfont2=\sevensy
  \textfont3=\tenex \scriptfont3=\tenex \scriptscriptfont3=\tenex
  \def\it{\fam\itfam\fourteenit}\textfont\itfam=\fourteenit
  \def\sl{\fam\slfam\fourteensl}\textfont\slfam=\fourteensl
  \def\bf{\fam\bffam\fourteenbf}\textfont\bffam=\fourteenbf%
  \scriptfont\bffam=\tenbf
  \scriptscriptfont\bffam=\sevenbf
  \normalbaselineskip=17pt
  \let\sc=\elevenrm
  \let\big=\tenbig
  \setbox\strutbox=\hbox{\vrule height8.5pt depth3.5pt width\z@}%
  \normalbaselines\rm}

\def\seventeenpoint{\def\rm{\fam0\seventeenrm}%
  \textfont0=\seventeenrm \scriptfont0=\fourteenrm \scriptscriptfont0=\tenrm
  \textfont1=\seventeeni \scriptfont1=\fourteeni \scriptscriptfont1=\teni
  \textfont2=\seventeensy \scriptfont2=\fourteensy \scriptscriptfont2=\tensy
  \textfont3=\tenex \scriptfont3=\tenex \scriptscriptfont3=\tenex
  \def\it{\fam\itfam\seventeenit}\textfont\itfam=\seventeenit
  \def\sl{\fam\slfam\seventeensl}\textfont\slfam=\seventeensl
  \def\bf{\fam\bffam\seventeenbf}\textfont\bffam=\seventeenbf%
  \scriptfont\bffam=\fourteenbf
  \scriptscriptfont\bffam=\twelvebf
  \normalbaselineskip=21pt
  \let\sc=\fourteenrm
  \let\big=\tenbig
  \setbox\strutbox=\hbox{\vrule height 12pt depth 6pt width\z@}%
  \normalbaselines\rm}

\def\ninepoint{\def\rm{\fam0\ninerm}%
  \textfont0=\ninerm \scriptfont0=\sixrm \scriptscriptfont0=\fiverm
  \textfont1=\ninei \scriptfont1=\sixi \scriptscriptfont1=\fivei
  \textfont2=\ninesy \scriptfont2=\sixsy \scriptscriptfont2=\fivesy
  \textfont3=\tenex \scriptfont3=\tenex \scriptscriptfont3=\tenex
  \def\it{\fam\itfam\nineit}\textfont\itfam=\nineit
  \def\sl{\fam\slfam\ninesl}\textfont\slfam=\ninesl
  \def\bf{\fam\bffam\ninebf}\textfont\bffam=\ninebf \scriptfont\bffam=\sixbf
  \scriptscriptfont\bffam=\fivebf
  \normalbaselineskip=11pt
  \let\sc=\sevenrm
  \let\big=\ninebig
  \setbox\strutbox=\hbox{\vrule height8pt depth3pt width\z@}%
  \normalbaselines\rm}

\def\eightpoint{\def\rm{\fam0\eightrm}%
  \textfont0=\eightrm \scriptfont0=\sixrm \scriptscriptfont0=\fiverm%
  \textfont1=\eighti \scriptfont1=\sixi \scriptscriptfont1=\fivei%
  \textfont2=\eightsy \scriptfont2=\sixsy \scriptscriptfont2=\fivesy%
  \textfont3=\tenex \scriptfont3=\tenex \scriptscriptfont3=\tenex%
  \def\it{\fam\itfam\eightit}\textfont\itfam=\eightit%
  \def\sl{\fam\slfam\eightsl}\textfont\slfam=\eightsl%
  \def\bf{\fam\bffam\eightbf}\textfont\bffam=\eightbf \scriptfont\bffam=\sixbf%
  \scriptscriptfont\bffam=\fivebf%
  \normalbaselineskip=9pt%
  \let\sc=\sixrm%
  \let\big=\eightbig%
  \setbox\strutbox=\hbox{\vrule height7pt depth2pt width\z@}%
  \normalbaselines\rm}

\def\tenbig#1{{\hbox{$\left#1\vbox to8.5pt{}\right.\n@space$}}}
\def\ninebig#1{{\hbox{$\textfont0=\tenrm\textfont2=\tensy
  \left#1\vbox to7.25pt{}\right.\n@space$}}}
\def\eightbig#1{{\hbox{$\textfont0=\ninerm\textfont2=\ninesy
  \left#1\vbox to6.5pt{}\right.\n@space$}}}

\def\footnote#1{\edef\@sf{\spacefactor\the\spacefactor}#1\@sf
      \insert\footins\bgroup\eightpoint
      \interlinepenalty100 \let\par=\endgraf
        \leftskip=\z@skip \rightskip=\z@skip
        \splittopskip=10pt plus 1pt minus 1pt \floatingpenalty=20000
        \smallskip\item{#1}\bgroup\strut\aftergroup\@foot\let\next}
\skip\footins=12pt plus 2pt minus 4pt 
\dimen\footins=30pc 

\newinsert\margin
\dimen\margin=\maxdimen
\def\titlefont{\seventeenpoint}
\loadtruetwelvepoint 
\loadtrueseventeenpoint

\def\eatOne#1{}
\def\ifundef#1{\expandafter\ifx%
\csname\expandafter\eatOne\string#1\endcsname\relax}
\def\notTrue{\iffalse}\def\isTrue{\iftrue}
\def\ifdef#1{{\ifundef#1%
\aftergroup\notTrue\else\aftergroup\isTrue\fi}}
\def\use#1{\ifundef#1\linemessage{Warning: \string#1 is undefined.}%
{\tt \string#1}\else#1\fi}


\global\newcount\refno \global\refno=1
\newwrite\rfile
\newlinechar=`\^^J
\def\@ref#1#2{\the\refno\n@ref#1{#2}}
\def\n@ref#1#2{\xdef#1{\the\refno}%
\ifnum\refno=1\immediate\openout\rfile=\jobname.refs\fi%
\immediate\write\rfile{\noexpand\item{[\noexpand#1]\ }#2.}%
\global\advance\refno by1}
\def\nref{\n@ref} 
\def\ref{\@ref}   
\def\lref#1#2{\the\refno\xdef#1{\the\refno}%
\ifnum\refno=1\immediate\openout\rfile=\jobname.refs\fi%
\immediate\write\rfile{\noexpand\item{[\noexpand#1]\ }#2\semi}%
\global\advance\refno by1}
\def\cref#1{\immediate\write\rfile{#1\semi}}

\def\preref#1#2{\gdef#1{\@ref#1{#2}}}

\def\semi{;\hfil\noexpand\break}

\def\listrefs{\vfill\eject\immediate\closeout\rfile
\centerline{{\bf References}}\bigskip\frenchspacing%
\input \jobname.refs\vfill\eject\nonfrenchspacing}

\def\inputAuxIfPresent#1{\immediate\openin1=#1
\ifeof1\message{No file \auxfileName; I'll create one.
}\else\closein1\relax\input\auxfileName\fi%
}
\def\NPB{Nucl.\ Phys.\ B}

\def\ZPC{Z.\ Phys.\ C}

\newif\ifWritingAuxFile
\newwrite\auxfile
\def\SetUpAuxFile{%
\xdef\auxfileName{\jobname.aux}%
\inputAuxIfPresent{\auxfileName}%
\WritingAuxFiletrue%
\immediate\openout\auxfile=\auxfileName}



\catcode`\@=\active
\catcode`@=12  
\catcode`\"=\active

\def\pol{\varepsilon}

\def\spa#1.#2{\left\langle#1\,#2\right\rangle}
\def\spb#1.#2{\left[#1\,#2\right]}
\def\lor#1.#2{\left(#1\,#2\right)}
\def\sand#1.#2.#3{%
\left\langle\smash{#1}{\vphantom1}^{-}\right|{#2}%
\left|\smash{#3}{\vphantom1}^{-}\right\rangle}
\def\sandp#1.#2.#3{%
\left\langle\smash{#1}{\vphantom1}^{-}\right|{#2}%
\left|\smash{#3}{\vphantom1}^{+}\right\rangle}
\def\sandpp#1.#2.#3{%
\left\langle\smash{#1}{\vphantom1}^{+}\right|{#2}%
\left|\smash{#3}{\vphantom1}^{+}\right\rangle}
\catcode`@=11  
\def\meqalign#1{\,\vcenter{\openup1\jot\m@th
   \ialign{\strut\hfil$\displaystyle{##}$ && $\displaystyle{{}##}$\hfil
             \crcr#1\crcr}}\,}
\catcode`@=12  

\newread\epsffilein    
\newif\ifepsffileok    
\newif\ifepsfbbfound   
\newif\ifepsfverbose   
\newdimen\epsfxsize    
\newdimen\epsfysize    
\newdimen\epsftsize    
\newdimen\epsfrsize    
\newdimen\epsftmp      
\newdimen\pspoints     
\pspoints=1bp          
\epsfxsize=0pt         
\epsfysize=0pt         
\def\epsfbox#1{\global\def\epsfllx{72}\global\def\epsflly{72}%
   \global\def\epsfurx{540}\global\def\epsfury{720}%
   \def\lbracket{[}\def\testit{#1}\ifx\testit\lbracket
   \let\next=\epsfgetlitbb\else\let\next=\epsfnormal\fi\next{#1}}%
\def\epsfgetlitbb#1#2 #3 #4 #5]#6{\epsfgrab #2 #3 #4 #5 .\\%
   \epsfsetgraph{#6}}%
\def\epsfnormal#1{\epsfgetbb{#1}\epsfsetgraph{#1}}%
\def\epsfgetbb#1{%
%
%
\openin\epsffilein=#1
\ifeof\epsffilein\errmessage{I couldn't open #1, will ignore it}\else
%
%
   {\epsffileoktrue \chardef\other=12
    \def\do##1{\catcode`##1=\other}\dospecials \catcode`\ =10
    \loop
       \read\epsffilein to \epsffileline
       \ifeof\epsffilein\epsffileokfalse\else
%
%
          \expandafter\epsfaux\epsffileline:. \\%
       \fi
   \ifepsffileok\repeat
   \ifepsfbbfound\else
    \ifepsfverbose\message{No bounding box comment in #1; using defaults}\fi\fi
   }\closein\epsffilein\fi}%
%
%
\def\epsfclipstring{}
\def\epsfsetgraph#1{%
   \epsfrsize=\epsfury\pspoints
   \advance\epsfrsize by-\epsflly\pspoints
   \epsftsize=\epsfurx\pspoints
   \advance\epsftsize by-\epsfllx\pspoints
%
%
   \epsfxsize\epsfsize\epsftsize\epsfrsize
   \ifnum\epsfxsize=0 \ifnum\epsfysize=0
      \epsfxsize=\epsftsize \epsfysize=\epsfrsize
      \epsfrsize=0pt
%
%
     \else\epsftmp=\epsftsize \divide\epsftmp\epsfrsize
       \epsfxsize=\epsfysize \multiply\epsfxsize\epsftmp
       \multiply\epsftmp\epsfrsize \advance\epsftsize-\epsftmp
       \epsftmp=\epsfysize
       \loop \advance\epsftsize\epsftsize \divide\epsftmp 2
       \ifnum\epsftmp>0
          \ifnum\epsftsize<\epsfrsize\else
             \advance\epsftsize-\epsfrsize \advance\epsfxsize\epsftmp \fi
       \repeat
       \epsfrsize=0pt
     \fi
   \else \ifnum\epsfysize=0
     \epsftmp=\epsfrsize \divide\epsftmp\epsftsize
     \epsfysize=\epsfxsize \multiply\epsfysize\epsftmp
     \multiply\epsftmp\epsftsize \advance\epsfrsize-\epsftmp
     \epsftmp=\epsfxsize
     \loop \advance\epsfrsize\epsfrsize \divide\epsftmp 2
     \ifnum\epsftmp>0
        \ifnum\epsfrsize<\epsftsize\else
           \advance\epsfrsize-\epsftsize \advance\epsfysize\epsftmp \fi
     \repeat
     \epsfrsize=0pt
    \else
     \epsfrsize=\epsfysize
    \fi
   \fi
%
%
   \ifepsfverbose\message{#1: width=\the\epsfxsize, height=\the\epsfysize}\fi
   \epsftmp=10\epsfxsize \divide\epsftmp\pspoints
   \vbox to\epsfysize{\vfil\hbox to\epsfxsize{%
      \ifnum\epsfrsize=0\relax
        \includegraphics{#1}%
      \else
        \epsfrsize=10\epsfysize \divide\epsfrsize\pspoints
        \includegraphics{#1}%
      \fi
      \hfil}}%
\global\epsfxsize=0pt\global\epsfysize=0pt}%
%
%
{\catcode`\%=12 \global\let\epsfpercent=
%
%
\long\def\epsfaux#1#2:#3\\{\ifx#1\epsfpercent
   \def\testit{#2}\ifx\testit\epsfbblit
      \epsfgrab #3 . . . \\%
      \epsffileokfalse
      \global\epsfbbfoundtrue
   \fi\else\ifx#1\par\else\epsffileokfalse\fi\fi}%
%
%
\def\epsfempty{}%
\def\epsfgrab #1 #2 #3 #4 #5\\{%
\global\def\epsfllx{#1}\ifx\epsfllx\epsfempty
      \epsfgrab #2 #3 #4 #5 .\\\else
   \global\def\epsflly{#2}%
   \global\def\epsfurx{#3}\global\def\epsfury{#4}\fi}%
%
%
\def\epsfsize#1#2{\epsfxsize}
%
%

\hfuzz=25pt

\def\dlips{d{\rm LIPS}}

\def\rg{r_\Gamma}
\def\Atree{A^{\rm tree}}
\def\Aloop{A^{\rm 1-loop}}

\def\eps{\epsilon}
\def\tr{{\rm tr}}
\def\Slash#1{\slash\hskip -0.17 cm #1}
\def\IntSet{{\cal F}}



\nopagenumbers
\loadfourteenpoint

\noindent

$\null$

\vskip -1.6 cm

hep-th/9512084

\rightline{SWAT-95-59}

\hfill November 1995

\def\ref{\nref}
\ref\GravityReview{
B.S. DeWitt, Phys.\ Rev.\ 162:1239 (1967)\semi
M. Veltman, in {\it Les Houches 1975,
Methods in Field Theory}, ed R. Balian and J. Zinn-Justin,
(North Holland, Amsterdam, 1976)}

\ref\GNV{ M.T.\ Grisaru, P.\  van Nieuwenhuizen and J.A.M. Vermaseren,
Phys.\ Rev.\ Lett.\ 37:1662 (1976)}

\ref\HVb{G. 't\ Hooft and M.\ Veltman,
Ann. Inst. Henri Poincar\'e 20:69 (1974)}

\ref\MatterEM{S. Deser and  P. van Nieuwenhuizen,
Phys.\ Rev.\ D10:411 (1974)\semi
M.T. Grisaru, P. van Nieuwenhuizen and  C.C.\ Wu,
Phys.\ Rev.\ D12:1813 (1975)}

\ref\Background{G. 't Hooft,
Acta Universitatis Wratislavensis no.\
38, 12th Winter School of Theoretical Physics in Karpacz; {\it
Functional and Probabilistic Methods in Quantum Field Theory},
Vol. 1 (1975)\semi
B.S.\ DeWitt, in {\it Quantum Gravity II}, eds. C. Isham, R.\ Penrose and
D.\ Sciama (Oxford, 1981)\semi
L.F.\ Abbott, Nucl.\ Phys.\ B185:189 (1981)\semi
L.F\ Abbott, M.T.\ Grisaru and R.K.\ Schaeffer,
Nucl.\ Phys. {B229}:372 (1983)}

\ref\GoroffSagnotti{ M.H. Goroff and A. Sagnotti,
Phys. Lett. 160B:81(1985), Nucl. Phys. B266:709 (1986)}

\ref\VanDVen{A.E.M.\ van de Ven,  Nucl.\ Phys.\ B378:309 (1992)}

\ref\StringBased{
Z. Bern and D.A.\ Kosower, Phys.\ Rev.\ Lett.\ 66:1669 (1991);
Nucl.\ Phys.\ B379:451 (1992) \semi
Z. Bern, L. Dixon and D.A. Kosower, Phys.\ Rev. Lett.\
70:2677 (1993); Nucl.\ Phys. B437:259 (1995)}

\ref\Gravity{Z. Bern, D.C. Dunbar and T. Shimada,
Phys.\ Lett.\ 312B:277 (1993)}

\ref\GravityString{D.C.\  Dunbar and P.S.\ Norridge,
Nucl.\ Phys.\ B433:181 (1995)}

\ref\Cutkosky{L.D.\ Landau, Nucl.\ Phys.\ 13:181 (1959)\semi
 S. Mandelstam, Phys.\ Rev.\ 112:1344 (1958), 115:1741 (1959)\semi
 R.E.\ Cutkosky, J.\ Math.\ Phys.\ 1:429 (1960)}

\ref\SusyFour{Z. Bern, D.C. Dunbar, L. Dixon and D.A. Kosower,
 Nucl.\ Phys.\ B425:217 (1994) }

\ref\SusyOne{Z. Bern, D.C. Dunbar, L. Dixon and D.A. Kosower,
Nucl.\ Phys.\ B435:59 (1995) }

\ref\BernMorgan{
Z.\ Bern and A.\ Morgan, hep-ph/9511336}

\ref\KST{Z. Kunszt, A. Signer and Z. Trocsanyi,
 Nucl.\ Phys.\ B420:550 (1994)}

\ref\LD{L. Dixon, private communication }

\ref\Weinberg{S. Weinberg, Phys. Rev.\ 140:B516 (1962) }

\ref\Mapping{Z. Bern and D.C.\ Dunbar,  Nucl.\ Phys.\ B379:562 (1992)}

\ref\FirstQ{
E.S.\ Fradkin and A.A.\ Tseytlin, Phys. Lett. 158B:316 (1985);
163B:123 (1985); Nucl. Phys. B261:1 (1985)\semi
M. Strassler, Nucl.\ Phys.\ {B385}:145 (1992)\semi
M.G. Schmidt and C. Schubert, Phys.\ Lett.\ 318B:438 (1993)\semi
D.G.C.\ McKeon, Ann. Phys. (N.Y.) 224:139 (1993)}

\ref\SusyReg{W. Siegel, Phys.\ Lett.\ 84B:193 (1979)\semi
D.M.\ Capper, D.R.T.\ Jones and P. van Nieuwenhuizen, Nucl.\ Phys.\
B167:479 (1980)\semi
L.V.\ Avdeev and A.A.\ Vladimirov, Nucl.\ Phys.\ B219:262 (1983)\semi
I.\ Jack, D.R.T.\ Jones and K.L. Roberts,  Z. Phys. C63:151 (1994)  }

\ref\Sannan{S.\ Sannan, Phys.\ Rev.\ D34:1748 (1986)}

\ref\PV{L.M.\ Brown and R.P.\ Feynman, Phys.\ Rev.\ 85:231 (1952)\semi
G.\ Passarino and M.\ Veltman, Nucl.\ Phys.\ {B160:151} (1979)\semi
G. 't Hooft and M. Veltman, \NPB{153:365 (1979)}\semi
R. G. Stuart, Comp.\ Phys.\ Comm.\ 48:367 (1988)\semi
R. G. Stuart and A. Gongora, Comp.\ Phys.\ Comm.\ 56:337 (1990)}

\ref\OtherMPoint{
D. B. Melrose, Il Nuovo Cimento 40A:181 (1965)\semi
W. van Neerven and J.A.M. Vermaseren, Phys.\ Lett.\ 137B:241 (1984)\semi
G. J. van Oldenborgh and J.A.M. Vermaseren, \ZPC{46:425 (1990)}\semi
G. J. van Oldenborgh, PhD thesis, University of Amsterdam (1990)\semi
A. Aeppli, PhD thesis, University of Zurich (1992)}

\ref\Factorise{Z.\ Bern and G.\ Chalmers,
Nucl.\ Phys.\ B447:465 (1995)}

\ref\Berends{F.A. Berends, W.T.\ Giele and H. Kuijf,
Phys. Lett.\ {211B}:91 (1988)}

\ref\SpinorGravity{
S.F.Novaes and D.Spehler,
Phys.Rev. D44:3990 (1991); Nucl.Phys.\ B371:618(1992)\semi
H.T.\ Cho, K.L.\ Ng, Phys.\ Rev.\ D47:1692 (1993) }

\ref\XZC{%
F.\ A.\ Berends, R.\ Kleiss, P.\ De Causmaecker, R.\ Gastmans and T.\ T.\ Wu,
        Phys.\ Lett.\ 103B:124 (1981)\semi
P.\ De Causmaeker, R.\ Gastmans,  W.\ Troost and  T.\ T.\ Wu,
Nucl. Phys. B206:53 (1982)\semi
R.\ Kleiss and W.\ J.\ Stirling,
   Nucl.\ Phys.\ B262:235 (1985)\semi
   J.\ F.\ Gunion and Z.\ Kunszt, Phys.\ Lett.\ 161B:333 (1985)\semi
 R.\ Gastmans and T.T.\ Wu,
{\it The Ubiquitous Photon: Helicity Method for QED and QCD} (Clarendon Press)
(1990)\semi
Z.\ Xu, D.-H.\ Zhang and L. Chang, Nucl.\ Phys.\ B291:392 (1987)}

\ref\Paul{ P.S.\ Norridge, in preparation }

\ref\Mahlon{G.D.\ Mahlon , Phys.\ Rev. D49 (1994) 2197}

\ref\Zak{M.T. Grisaru and J. Zak, Phys. Lett.\ {90B}:237 (1980)}

\ref\Susy{ M.T. Grisaru, H.N. Pendleton and P.  van Nieuwenhuizen,
Phys. Rev. {D15}:996 (1977)\semi
M.L.\ Mangano and S.J. Parke, Phys.\ Rep.\ {200}:301 (1991)}

\vskip 1in
{\titlefont\centerline{\bf Infinities within Graviton}}
{\titlefont\centerline{\bf Scattering Amplitudes}}
\vskip .3cm

\vskip .5 in


\centerline{
{\bf David C. Dunbar}\footnote{${}^\dagger$}{d.c.dunbar@swan.ac.uk}
and {\bf Paul S. Norridge}\footnote{${}^\ddagger$}{p.s.norridge@swan.ac.uk}
}

\vskip 0.7 truecm \baselineskip12pt
\centerline{\it Department of Physics}
\centerline{\it University of Wales, Swansea}
\centerline{\it Swansea, SA2 8PP}
\centerline{\it UK }

\vskip 1.2 truecm \baselineskip12pt

\centerline{\bf Abstract } { \narrower\smallskip \smallskip
We present unitarity as a method for determining the
infinities present in graviton scattering amplitudes.
B
The infinities are a combination of IR and UV.
By understanding
the soft singularities we may
extract the UV
infinities and
relate these to counter-terms in the effective action.
As an demonstration of this method
we rederive the UV infinities present
at one-loop when gravity is coupled to matter.}

\baselineskip14pt

\vglue 0.3cm

\vfill\eject

\footline={\hss\tenrm\folio\hss}

\section{ Introduction}

Calculations in perturbative gravity [\use\GravityReview] are
notoriously difficult to perform.  In particular determining the
renormalisability of gravity whether coupled to matter or not is a
difficult issue.  General arguments regarding the symmetries in the
action may allow or prohibit counter terms.  Such arguments show that
the UV infinities vanish on-shell up to one-loop for pure gravity
amplitudes and for up to two-loops for the particular matter coupling
in supergravities [\use\GNV].  However in gravity coupled to general
matter, there are possible counter terms at one-loop level and a
calculation must be performed to determine the coefficient of the
potential counter-term.  Such calculations have been done for gravity
coupled to matter and the explicit coefficients obtained
[\use\HVb,\use\MatterEM,\use\GNV].  These calculations use an algorithm
due to `t Hooft and Veltman [\use\HVb] which examine counter terms in
the effective action using the background field method
[\use\Background].  (The coefficients of the counter terms in pure
gravity at two loops have been calculated
[\use\GoroffSagnotti,\use\VanDVen] with non-zero result.)

Recently developments have been made in the calculation of on-shell
amplitudes using string inspired techniques and the Cutkosky rules.
These have enabled new results to be obtained both in QCD
[\use\StringBased] and perturbative quantum gravity
[\use\Gravity,\use\GravityString].  In this paper we show how these
techniques may be used to detect the effect of the counter terms in
the one-loop amplitudes - and in fact extract their coefficient. This
calculational method is quite distinct from usual techniques.

The Cutkosky rules [\use\Cutkosky] relate two amplitudes 'sewn'
together to the discontinuous parts of an amplitude at higher-order.
Knowing the lower order expressions we may impose constraints upon the
higher order parts of the amplitude. (For example, the Cutkosky rules
relate the imaginary parts of one-loop amplitudes to tree amplitudes
sewn together.)  The constraints these rules impose upon amplitude can
completely reconstruct some amplitudes and constrain the expressions
for others [\use\SusyFour,\use\SusyOne].  In gravity this technique
has provided a check for the 1-loop results from the string-inspired
rules although several of the four point amplitudes could have been
reconstructed entirely from the cuts.  In general it is the ``more
supersymmetric'' amplitudes which are cut-constructable.  For general
amplitudes we may construct, via the Cutkosky rules, expressions
containing the correct cuts in all channels.  Although such expressions
contain finite ambiguities, we may use these expressions to evaluate the
infinities present in an amplitude.  In fact, we will demonstrate the
one-loop UV-infinities present in matter-coupled gravity using this
method.  To enable us to disentangle the UV and IR infinities we also
determine the general structure of the one-loop soft infinities.

Finally, although amplitudes naively calculated using the Cutkosky
rules contain finite ambiguities these may be resolved by a more
sophisticated use of the rules [\use\BernMorgan].  We illustrate this
with a specific calculation at one-loop involving graviton scattering
in a theory with massive scalar matter.

\section{Infrared divergences}

As we will see later, one can, by using unitarity, extract the
infinities present in an amplitude.  However, this method will not
distinguish between UV and IR infinities. Nonetheless, by knowing the
expected form of the IR singularities we may identify the remaining UV
infinities. \footnote{${}^\dagger$}{We thank Lance Dixon [\use\LD]
for help in realising this possability}

In this section we determine the IR singularities present in
amplitudes involving massless particles.  The IR singularities in a
QCD amplitude have been calculated [\use\KST] and our determination of
the IR singularities for gravity follow in a similar manner.
One may also deduce the form of the infra-red divergences using general
arguments to imply universality  and then extract the form
from a specific amplitude [\use\LD].
As is
well known [\use\Weinberg] IR singularities in a on-shell loop amplitude
occur in soft limits of the loop momentum integrals.
That is, when the momentum of internal
propagators goes to zero.  However, not all such occurances produce IR
singularities. Any massless propagator through which loop momentum
flows will be singular for a specific value of the loop momentum but
this does not usually yield a singularity. To see this note that, in
four dimensions, the singular part of the momentum integral can be
expressed
$$
\int d^4 p {1\over p^2} \sim \int_0 |p|^3 d|p| {1\over |p|^2}
\anoneqn
$$
which is finite at the lower range of integration, $|p| \rightarrow 0$
the singularity having been suppressed by the factor of $|p|^3$.  (We
always choose gauges where the propagator is Feynman-like $\sim
1/p^2$. Gauges closely related to the string based rules
[\use\Mapping] and the ``World-line'' approach [\use\FirstQ] have this
feature.)  In general three adjacent propagators must vanish
simultaneously to obtain a soft divergence. To see this note that
three adjacent propagators in a loop will be of the form (we shift the
loop momenta such that the middle momenta is the integration momenta)
$$
{1\over (p+K_1)^2} {1\over p^2} {1\over (p-K_2)^2}
\sim
{1 \over (p^2 +2p\cdot K_1+ K_1^2)}
{1\over p^2}
{ 1 \over (p^2 -2p\cdot K_2 +K_2^2) }
\eqn\ThreeProps
$$
We may obtain an IR singularity provided $K_1^2=0$ and $K_2^2=0$.  One can
imagine more complicated situations in multi-loop diagrams where
adjacent propagators vanish at choices of the multiple
integrations. However the equivalent factors to $|p|^3$ in the above
equation are of higher powers and suppress the infinities more
strongly.  After inspection of the possibilities we find the above
situation is the generic case for a soft divergence in the
loop-momentum integral.

We will only, for general kinematics, find propagators where $K_1^2=0$
and $K_2^2=0$ if $K_1$ and $K_2$ are adjacent external on-shell
momentum (for massless particles.)  The middle propagator $1/p^2$ is
then joining together the two external legs.  One may see this general
arrangement in
\fig\SoftFigure .
Thus the soft limit may be found by taking the amplitude with one less
loop and adding a soft particle between two of the external lines.
Consider first an individual diagram, rather than the whole amplitude.
We initially restrict ourselves to a diagram containing only
gravitons.  Throughout, amplitudes will be calculated using
dimensional reduction [\use\SusyReg].

Consider adding a soft graviton to an graviton scattering diagram
by attaching a graviton between external legs with momentum $k_1$ and
$k_2$ as in fig.~\use\SoftFigure .
In doing so we must add a propagator for the graviton, two three
point vertices and
two extra propagators with momenta $k_1+p$ and $k_2-p$ to the normal
Feynman
diagram expression.
The soft singularity, as in eqn.~(\use\ThreeProps),
will occur in that region of the $\int d^D p$
integral where $p$ is close to zero.

\LoadFigure\SoftFigure{\baselineskip 13 pt
\noindent\narrower\ninerm Representation of the soft limit calculation.}
{\epsfysize 1.5in}
{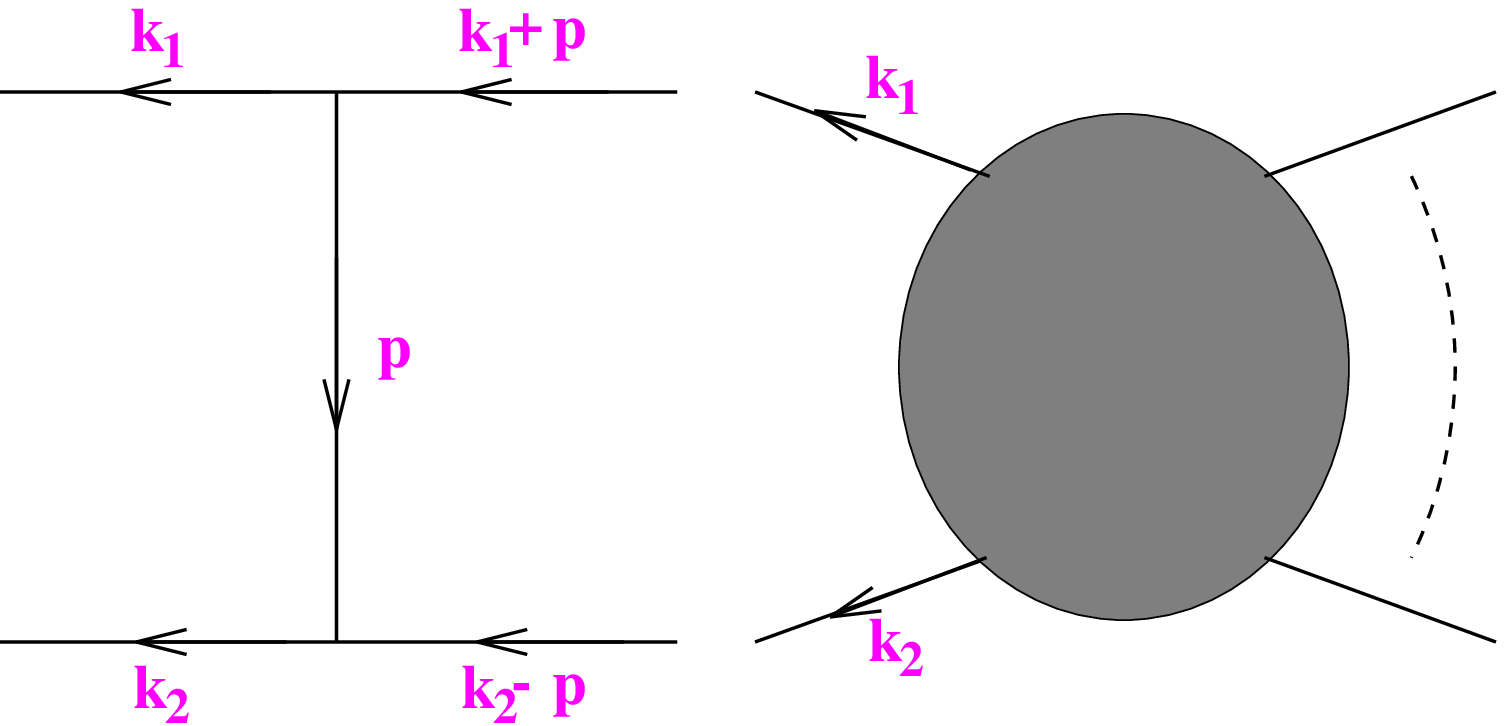}

Initially consider the soft-divergences in a one-loop diagram. Thus
the soft graviton leg will connect two external legs of a tree
diagram. Take the off-shell tree diagram as a function of the two
external momentum $k_1$ and $k_2$ only and of the indices on these two
legs only (keeping all others fixed),
$A^{\rm tree }_{(\alpha\beta),(\gamma\mu)}(k_1,k_2)$
The on-shell tree diagram is
the contraction of this with the external polarisation tensors
(together with $k_1$ and $k_2$ becoming onshell),
$$
{\epsilon_1}^{\alpha\beta}
{\epsilon_2}^{\gamma\mu}\, A^{\rm
tree}_{(\alpha\beta),(\gamma\mu)}(k_1,k_2)
\anoneqn
$$
the amplitude being the sum of all such diagrams.
The $1-$loop diagram can be written in terms of the off-shell
tree diagram with the additional three propagators and
two vertices,
$$
\int {d^{4-2\epsilon}p \over (2\pi)^{4-2\eps} }
{ T^{(\alpha\beta),(\mu\nu)}(k_1,k_2,p)\over
(p+k_1)^2 p^2 (p-k_2)^2 }\,
A^{\rm tree}_{(\alpha\beta),(\mu\nu)}(k_1+p,k_2-p)
\anoneqn
$$
The tensor $T$ is
$$
\eqalign{
T^{(\alpha\beta),(\mu\nu)}(k_1,k_2,p)
&={\epsilon_1}^{\alpha_1\beta_1}
{\epsilon_2}^{\mu_1\nu_1}
 V_{(\alpha_1\beta_1),(\delta\gamma),(\alpha\beta)}(k_2,-p,-k_2+p)\cr
&\qquad\qquad\times
P_{(\delta\gamma),(\sigma\rho) }\;
V_{(\mu_1\nu_1),(\sigma\rho),(\mu\nu)}(k_1,p,-k_1-p)
}
\anoneqn
$$
where
$V_{(\mu\nu),(\sigma\rho),(\gamma\eta)}$  is the 3-graviton vertex
and $P_{(\alpha\beta),(\sigma\rho) }$ is the propagator
[\use\GravityReview,\use\Sannan,\use\VanDVen].
In the $p\rightarrow0$ limit, we find $T$ becomes
$$
T^{(\alpha\beta),(\gamma\mu)}(k_1,k_2,p)
=
-{\kappa^2 \over 4} \,{\epsilon_1}^{{\alpha\beta}}{\epsilon_2}^{{\gamma
\mu}}
(k_1+ k_2)^{4}
+O(p)
\eqn\TensorLimit
$$
To examine the leading soft singularity we must therefore
look at
the integral,
$$
\int{ d^{4-2\epsilon}p \over (2\pi)^{4-2\eps } }
{ 1 \over (p+k_1)^2 p^2 (p-k_2)^2 }
\anoneqn
$$
which can be evaluated by usual Feynman parameter methods with the result,
$$
{ -i \rg  \over (4\pi)^{2-\epsilon }(-s_{12} )^{1+\epsilon} }
{ 1 \over \epsilon^2}
\anoneqn
$$
where
$\rg=\Gamma^2(1-\epsilon)\Gamma(1+\epsilon)/\Gamma(2-\epsilon)$
and $s_{12}=(k_1+k_2)^2$.
\footnote{${}^\dagger$}{We shall always give amplitudes (and integrals)
in the unphysical region where all momentum invariants such as $s_{ij}$
are negative. One may continue back to the physical regime by
$\ln (-s) \rightarrow \ln(|s|) -i\pi \Theta(s)$ etc.
We also use the convention that all particles are outgoing.}
The leading singularity as $p\rightarrow 0$ is then simply a factor
multiplying the tree diagram. Summing over all diagrams then gives
a factor multiplying the tree amplitude. To be precise,
we find that the IR divergence, due to a soft graviton
exchange between external
legs $1$ and $2$, in a one-loop graviton scattering amplitude is
$$
{ i\rg  \over (4\pi)^{2-\epsilon} }
{ \kappa^2 \over 4 \eps^2 } \times
 (-s_{12} )^{1-\epsilon}  \times A^{\rm tree}
\anoneqn
$$
There are a few subtleties with the above result. Firstly, we have to
take the limit of an off-shell amplitude $A^{\rm tree}(k_1+p, k_2-p)$
as $p\rightarrow 0$. This is trivial for a tree amplitude but for the
$n$-loop case is more subtle. In general this will yield the on-shell
result for an amplitude although this is not true merely for a single
$n$-loop diagram [\use\Factorise].  Secondly there are sub-leading
terms in eqn.(\use\TensorLimit).  These vanish on summing over diagrams
because we are then dealing with a physical on-shell amplitude.

We have calculated the IR divergence due to soft graviton exchange
between a specific pair of external legs.  The total IR divergence in
a one-loop amplitude is simply the above result summed over all pairs
of external legs.
$$
{i r_\Gamma\over(4\pi)^{2-\epsilon}}
{\kappa^2   \over 4\epsilon^2} \;\times  A^{\rm tree}
\times \sum_{i\neq j}^m (-s_{ij} )^{1-\epsilon}
\eqn\SoftAns
$$
In gauge theories, amplitudes contain IR divergences from the
self-energy corrections to the external legs.  [\use\KST].  However,
by power counting, such divergences do not occur for external graviton
legs.

Consider a specific example:
namely that of a 1-loop four graviton amplitude. These are given in
ref.~[\use\GravityString]. Also let us look at the specific helicity
configuration
$\Aloop(1^-,2^-,3^+,4^+)$.
Summing over all pairs of legs the expected IR divergence is
$$
2 \big((-s)^{1-\epsilon}+(-t)^{1-\epsilon}+(-u)^{1-\epsilon}\big)
\times{i r_\Gamma\over(4\pi)^{2-\epsilon}}
 {\kappa^2  \over 4\epsilon^2} A^{\rm tree}(1^-,2^-,3^+,4^+)
\anoneqn
$$
(A factor of $2$ arises because $s_{12}=s_{34}=s$.)
This is
$$
\eqalign{
{i r_\Gamma \kappa^2 \over(4\pi)^{2-\epsilon}}
\bigg(
{\big(s \ln(-s) + t \ln(-t) + u \ln(-u)\big)\over 2\epsilon}
\bigg)A^{\rm tree}(1^-,2^-,3^+,4^+)
}
\anoneqn
$$
For this amplitude we expect  no
UV singularities from the general result that
pure gravity is one-loop UV finite.
If we examine the complete result [\use\GravityString],
$$
\eqalign{
\Aloop(1^-,2^-,&3^+,4^+) =
{istu\kappa^2r_{\Gamma}\over 4(4\pi)^{2-\eps}}
{\Atree(1^-,2^-,3^+,4^+)}
\cr&\times\biggl(
 {2 \over \epsilon } \bigg(  {\ln(-u) \over st} + {\ln(-t) \over su}
+  {\ln(-s) \over tu}  \biggr)
+\hbox{finite terms}\biggr)
}\anoneqn
$$
we find the expected IR infinity structure.

In amplitudes with scalars and gravitons, there are also soft
divergences which  can be calculated similarly. For
the case of a scalars and gravitons there are six configurations which
must be considered, as shown in
\fig\SoftScalarsFigure .
In this figure scalars lines are dashed lines whereas graviton
lines are solid.  The analysis in all cases closely follows the pure
gravity calculation.  We find that (a), (b) and (c) all give the same
result as the pure gravity case, (\use\SoftAns); (d) and (e) give no
soft contribution.  The case (a), (b) and (c) are the diagrams where a
soft graviton is exchanged between the two external legs.  The result
in eqn.~(\use\SoftAns) is then universal. That is, true whether
external legs are gravitons or scalars.
We can also consider the IR divergences with other types of external
particle and find the same universal nature of the
soft divergences.
This universality may be expected upon
general arguments [\use\LD].

\LoadFigure\SoftScalarsFigure{\baselineskip 13 pt
\noindent\narrower\ninerm
Possible soft contributions to graviton-scalar amplitudes:
}
{\epsfysize 2.0in}
{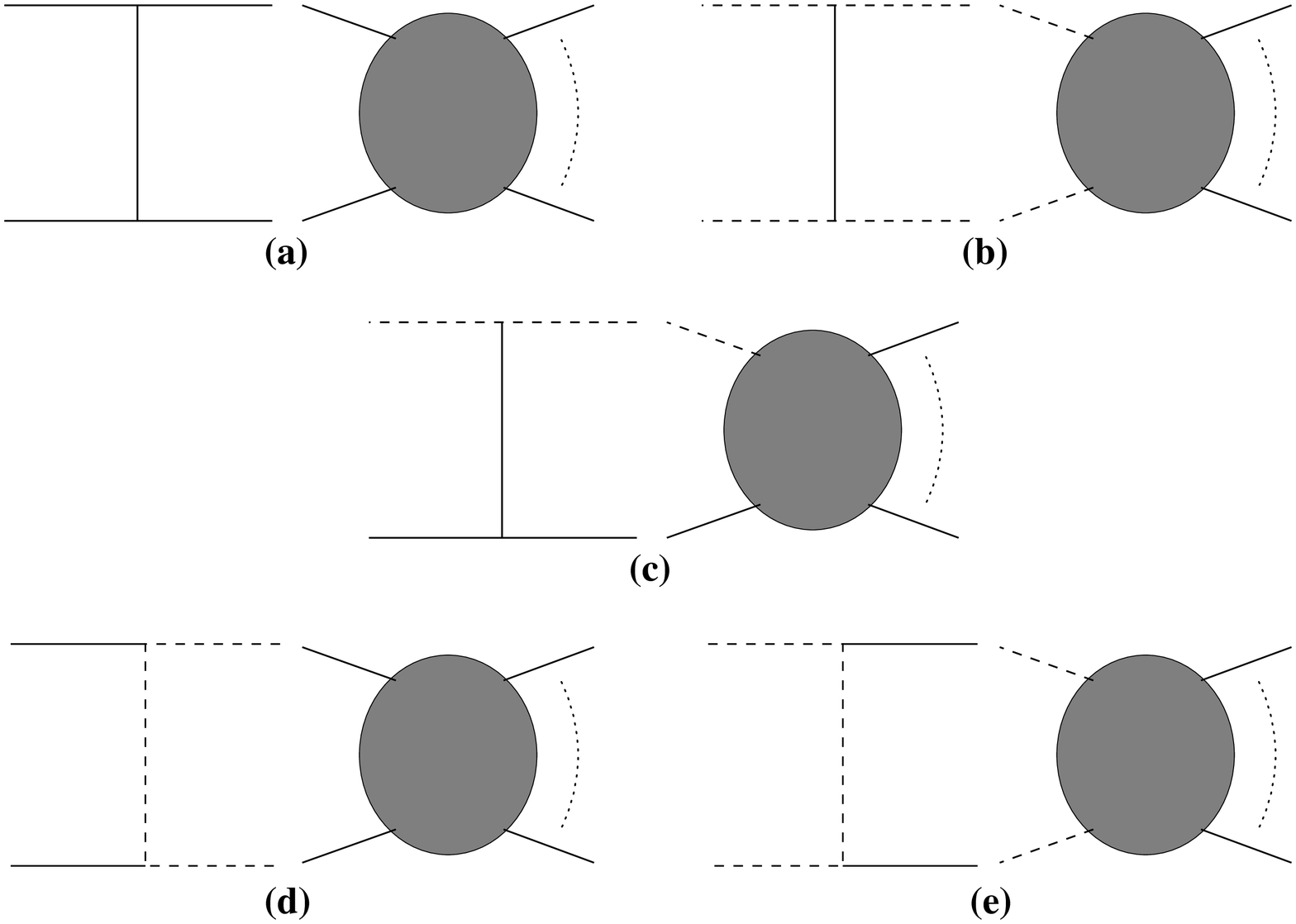}

\section{Divergences from the Cutkosky Rules}

As is well known, [\use\Cutkosky], unitarity in the
form of the Cutkosky rules,
relates the imaginary part of a one-loop amplitude to the product of
two tree amplitudes integrated over all
intermediate states.
In ref.[\use\SusyOne] situations where the Cutkosky rules
determine the amplitude entirely (and not merely the imaginary
part ) were investigated with the result that in supersymmetric gauge
theories the amplitudes may be determined by the Cutkosky rules alone.
For gravity only a few amplitudes may be determined from unitarity alone
[\use\StringBased]. However, we may determine the infinity structure
entirely from the Cutkosky rules.

In general, we consider the cut in the channel
$(k_{m_1}+k_{m_1+1}+\cdots+k_{m_2-1}+k_{m_2})^2$ for the
$1$-loop amplitude $A(1,2,\ldots,n)$, depicted in
\fig\CutsFigure\ and given by
$$
\eqalign{{i \over 2}
  &\int \dlips(-\ell_1,\ell_2)
  \ A^{\rm tree}(-\ell_1,m_1,\ldots,m_2,\ell_2)
  \ A^{\rm tree}(-\ell_2,m_2+1,\ldots,m_1-1,\ell_1). \cr
}
\eqn\CutEquation
$$

\LoadFigure\CutsFigure{\baselineskip 13 pt
\noindent\narrower\ninerm
The cut to be evaluated in eqn.(\CutEquation)
}
{\epsfysize 2.0in}
{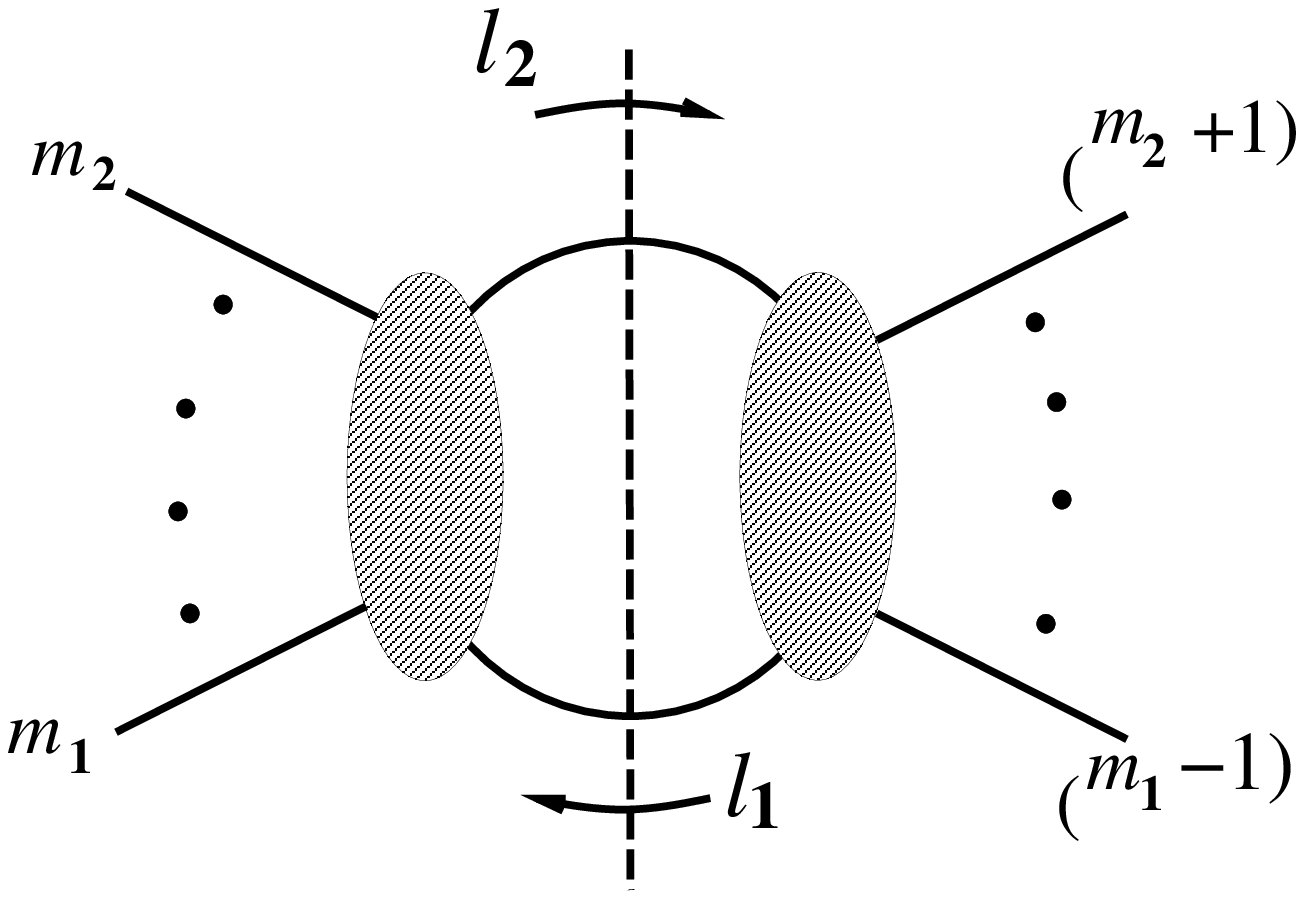}

In a $1$-loop amplitude with massless particles, the form of the
integrals which may appear is well known
( see for example refs.~[\use\PV,\use\OtherMPoint])
and the one-loop amplitude can be written in the form,
$$
  A_n =\  \sum_{i|I_i\in \IntSet_n} c_i I_i\ ,
\eqn\AnDecomp
$$
where the coefficients $c_i$ are {\it rational\/} functions of the
momentum invariants and $\IntSet_n$ is a set of integral functions.
In the appendices of [\use\SusyOne] a particular choice of $\IntSet_n$
is given although many are possible [\use\PV,\use\OtherMPoint].  All
the cuts arise from the integral functions $I_i$. These are for a
massless theory, rather simple, just arising from the cuts in
logarithms or the cuts in dilogarithms. One may see this by simple
inspection of the basis of functions in [\use\SusyOne].  In general
one cannot reconstruct the total amplitude from the simplistic
application of eqn.~(\use\CutEquation) however one may be able to
reconstruct the structure of the infinities. To see this consider one
of the integrals in the set $\IntSet_n$, The ``one-mass'' triangle
which depends only on the momentum invariant of the massive leg,
$s=K^2\neq 0$,
$$
I_{3;i}^{1\rm m} = {\rg\over\eps^2} (-s)^{-1-\eps}
={ \rg\over (-s) }\Bigl(  {1 \over\eps^2} -{ \ln(-s)\over\eps}
+{\ln^2(-s)\over 2} \Bigr)
+O(\eps)
\eqn\OneMassTriangle
$$
Clearly, the $1/ \eps^2$ pole is closely related to the
$\ln(-s)/\eps$ term. The latter is detectable from the Cutkosky rules
and from it we can reconstruct the first term. This is true in general,
by evaluating the cuts to order $\eps^0$ we can obtain the
divergences (or rather non-cut terms ) to order $\eps^{-1}$.
We will use this technique to obtain the divergences present in several
amplitudes.

As a technical issue we shall not evaluate (\use\CutEquation)
but instead evaluate the
the off-shell integral
$$
\eqalign{
  &{i\over 2} \int {d^{D} \ell_1 \over (2\pi)^D}
  \ A^{\rm tree}(-\ell_1,m_1,\ldots,m_2,\ell_2) {1\over \ell_2^2}
  \ A^{\rm tree}(-\ell_2,m_2+1,\ldots,m_1-1,\ell_1) {1\over \ell_1^2}
   \biggr|_{\rm cut} . \cr
}\eqn\caseacut
$$
whose cut in this channel is (\use\CutEquation) [\use\SusyFour].
This replacement is valid only in this channel. In evaluating this
off-shell integral, we may substitute
$\ell_1^2 = \ell_2^2 = 0$ in the numerator; terms with $\ell_1^2$
or $\ell_2^2$ in the numerator do not produce a cut in this
channel because the $\ell_1^2$ or $\ell_2^2$ cancels a cut propagator.
We emphasise that the cuts are evaluated
not for a lone diagram at a time, but for the whole amplitude.
After performing the cuts in all channels we may reconstruct the
infinities. In practice this means evaluation the
amplitude up to finite rational terms.

To illustrate this consider the case of a four-point function
with massless particles. The set of integrals $\IntSet_4$
can be chosen
to be fairly simple.
Firstly the ``scalar box integral''
$$
\eqalign{
I^{}_4 (s,t)\
&=\ \rg \, {1\over s t}
\biggl\{
{2 \over \eps^2} \Bigl[ ( -s)^{-\eps}+ (-t)^{-\eps} \Bigr]
- \ln^2 ( {-s \over - t})  - \pi^2 \biggr\} \ ,
\cr
&=\ \rg \, {1\over s t}
\biggl\{ {4\over \eps^2} -{ 2\ln( -s)+2\ln(-t)\over \eps}
+2\ln( -s)\ln(-t)  - \pi^2 \biggr\}
\cr}
\eqn\ZeroMassBox
$$
together with $I^{}_4 (s,u)$ and $I^{}_4 (t,u)$.
Then the ``triangle and bubble integrals''.
$$
\eqalign{
I_{3}(s)  &= {\rg\over\eps^2} (-s)^{-1-\eps}
= -{\rg\over s } \Bigl( {1 \over \eps^2 } - { \ln(-s) \over \eps}
+{ \ln^2(-s) \over 2} \Bigr)
\cr
I_2(s) &={\rg\over\eps (1-2\eps) } (-s)^{-\eps} =
\rg \Bigl( {1\over \eps } - \ln(-s) + 2 \Bigr)
\cr
J_2(s) & =\rg
\cr
}
\eqn\bubbles
$$
and the corresponding functions of $u$ and $t$.
The function $J_2(s)$ is a linear combination of bubble integrals
(see section~3 of ref~[\use\SusyOne]. ).  In supersymmetric amplitudes
the set $\IntSet_4$ may be more restrictive. (Specifically
$J_2$ may be
absent.)
A general massless four-point amplitude may be written thus,
$$
\eqalign{ A_4 =&
a_1 I^{}_4 (s,t) \; +a_2 I^{}_4 (s,u) \; +a_3 I^{}_4 (t,u)
\; +b_1 I_{3}(s) \; +b_2 I_{3}(t) \; + b_3I_{3}(u)
\cr
\; & +c_1 I_2(s)  \; +c_2 I_2(t) + \; c_3I_2(u) \; +d J_2
\cr}
\anoneqn
$$
where the coefficients $a_i$ etc. are rational
functions of the momentum
invariants.
\footnote{${}^\dagger$ }{If we were to consider the amplitude to higher
order than $O(\eps^0)$ we would have to expand our basis to include
further integral functions.}
With this choice of $\IntSet_4$ it is clear which coefficients may be
fixed by the cuts. For example the only integral function
containing a $\ln(-s)\ln(-t)$ term is $I^{}_4 (s,t)$ so by examining
the $\ln(-s)$ cut we can pick out $a_1$ from the $\ln(-t)$ term in the
cut. (Or alternatively the $\ln(-s)\ln(-t)$ term if we evaluate
eqn.~(\caseacut ).) Similarly by examining the other channels the $a_i$ are
simply fixed. The remaining
$\ln^2(-s)$ contribution fixes  $b_1$ and similarly the other $b_i$.
Finally the $\ln(-s)$ term will
determine $c_1$. So the $a_i$, $b_i$ and $c_i$ may be determined from
the cuts with the only remaining ambiguity arising in the $d J_2$ term.
For our purposes this shows that the only ambiguity will be in finite
rational  terms - and thus we may determine the infinity structure
purely from the cuts.

\section{Example: Gravity Coupled to Scalar Matter}

In this section we will show how the results of the previous sections
may be used to
determine the UV divergences in a specific example.
The example we choose is that of gravity coupled to
scalar matter. The form of the UV divergences in this theory is
well known [\use\HVb]. Although, pure gravity is one-loop UV finite,
in the presence of scalar matter infinities are generated in amplitudes
which necessitate a counter-term in the action,
$$
{203 \over 320\eps } ( D_{\mu} \phi D^{\mu} \phi )^2
\eqn\ScalarCounter
$$
Such a term is not present in the original theory
and indicates that gravity coupled to scalar matter is non-renormalisable.
We hope to arrive at the same conclusions by a consideration
of amplitudes. The simplest amplitude effected by such a term is
one with four external scalars ( and no external gravitons).
We look at a theory with only one type of scalar.
The intermediate states in eqn.(\use\caseacut)
can be either gravitons or scalars. We
consider the easier case of intermediate
Scalars first. The tree amplitude for
four scalars (all the same flavour) is :
$$
\Atree (\phi_1,\phi_2,\phi_3,\phi_4) =
{i\kappa^2\over2}\left({1 \over s} \Bigl( t^2 +u^2 \Bigr)
+{1 \over t} \Bigl( s^2 +u^2
\Bigr)
+{1 \over u} \Bigl( s^2 +t^2 \Bigr)\right)
\anoneqn
$$
\def\DP#1#2{k_{#1}\cdot \ell_{#2}}
so the product of tree amplitudes appearing in the $s$-channel
cut is
$$ \eqalign{
\Atree(\phi_1,\phi_2,  \phi_{\ell_1}& , \phi_{\ell_2} )\,
\Atree (\phi_{\ell_2} , \phi_{\ell_1} , \phi_3,\phi_4 )
 \cr
&={\kappa^4\over4}\biggl(
{1 \over s} \Bigl(4(\DP21)^2 +
4(\DP11)^2
\Bigr)
+{1
\over 2(\DP21)}
\Bigl( s^2+ 4(\DP11 )^2\Bigr) \Bigr)
\cr
\null& \hskip 8.0 truecm
+{1 \over 2(\DP11)} \Bigl(
4(\DP21)^2 + s^2\Bigr)\biggr)
\cr&\qquad\times
\biggl(
{1 \over s}
\Bigl(4(\DP31)^2 + 4(\DP41)^2 \Bigr)
+{1
\over 2(\DP31)} \Bigl( s^2+ 4(\DP41)^2\Bigr)
\cr
\null& \hskip 8.0 truecm
+{1 \over 2(\DP41)} \Bigl(
4(\DP31)^2 + s^2\Bigr)\biggr)
\cr}
\anoneqn
$$
Inserting this into eqn.(\use\caseacut)
yields a variety of terms. For example the first term in the expansion
of the above has only two propagators ($1/\ell_1^2$ and $1/\ell_2^2$)
and so may be identified as a tensor bubble integral and hence evaluated.
In total the expansion contains boxes, triangles, and bubbles.
These may be evaluated fairly easily using standard techniques.
giving an expression for the logarithmic
parts
$$
\eqalign{
&{\kappa^4{r_\Gamma}\over(4\pi)^{2-\epsilon}}
\bigg(
{\left (u^{4}+2\,tu^{3}+3\,u^{2}t^{2}+2\,ut^{3}+t^{4}\right )\over 2ut}
\,\ln (-s){1\over\epsilon}\cr
&\qquad +{1\over4}\,\left (3\,u
t+2\,u^{2}+2\,t^{2}\right )\,\ln (-s)^{2}+{ {s^{3}
\rg\,\ln (-s)\ln (-t)}\over{2\,t}}+{{s^{3}{\rg}\,\ln (-s)\ln
(-u)}\over{2\,u}}\cr
&\qquad\qquad+{{1\over{240} }  {\left (-161\,t^{2}+39\,ut-161\,u^{2}
\right )
\ln (-s)}}
\bigg)
\cr}
\anoneqn
$$

Next consider the
contribution to the cut
from intermediate gravitons.
The tree-amplitudes  needed are those involving two scalars and two
gravitons,
$$
\eqalign{
&\Atree(\phi_1 , 2^+ , 3^+ , \phi_4 ) = 0
\cr
&\Atree(\phi_1 , 2^- , 3^+ , \phi_4 ) =
{\kappa^2 \spa1.2^4\spa4.2^4 \over (\spa1.2\spa2.3\spa3.4\spa4.1)^2 }
{st \over 4u}
\cr}
\eqn\TreeAmp
$$
where we have used a spinor helicity convention.
The spinor helicity method for vectors is an
explicit realisation of the polarisation tensors in terms of
spinors
$$
\pol^{(+)}_\mu (k;q) =
{\sand{q}.{\gamma_\mu}.k
\over \sqrt2 \langle q^- | k^+ \rangle },\hskip 1cm
\pol^{(-)}_\mu (k;q) =
{\sandpp{q}.{\gamma_\mu}.k
\over \sqrt{2} \langle k^+ | q^- \rangle},
\anoneqn
$$
where $|k^{\pm} \rangle $ is a Weyl spinor, with plus and minus helicities,
$k$ is the on-shell momentum of the vector and
$q$ is an arbitrary reference momentum satisfying $q^2 =0$,
$k\cdot q \not = 0$.
The spinor helicity method for
gravitons [\use\Berends,\use\SpinorGravity] is related to that for
vectors
[\use\XZC] by
$$
\pol^{++}_{\mu\nu} = \pol^+_{\mu} \bar{\pol}^+_{\nu} ,
\hskip 2cm \pol^{--}_{\mu\nu} = \pol^-_{\mu} \bar{\pol}^-_{\nu}
\anoneqn
$$
where $\pol^{\pm\pm}$ are the graviton helicity polarisations and
$\pol^{\pm}$ are the vector helicity polarisations defined by Xu,
Zhang and Chang.
We use the notation for spinor inner products
$\langle k_1^- | k_2^+ \rangle = \langle 1 2 \rangle$ and
$\langle k_1^+ | k_2^- \rangle = \spb1.2 $. The use of spinor helicity
techniques has proved extremely useful in QCD calculation
and we will take advantage of the benefits here also.
All states are taken to be outgoing and may have plus or minus helicity.

Equation  (\use\TreeAmp) implies that the only
contribution comes from cuts with
gravitons of differing helicity across the cut.
The contribution to the cut from these will be
$$
\eqalign{
\Atree(\phi_1& , \phi_2 , \ell^-_1 , \ell^+_2 )\,
\Atree(\phi_3 , \phi_4 , \ell^-_2 ,
\ell^+_1)    \cr
 &={\kappa^4\over16}
{ \spa1.{\ell_1}^4\spa2.{\ell_1}^4 \over
(\spa1.{\ell_1}\spa{\ell_1}.{\ell_2}\spa{\ell_2}.2\spa2.1)^2 }
{s(k_1 \cdot \ell_1) \over ({k_1}\cdot \ell_2)}
\times {\spa4.{\ell_2}^4\spa3.{\ell_2}^4 \over
(\spa3.{\ell_2}\spa{\ell_2}.{\ell_1}\spa{\ell_1}.4\spa4.3)^2 } {s(l_2\cdot k_3)
\over (\ell_1\cdot k_3)}
\cr
&= {\kappa^4 \over 16\,s^2}{
(tr_-(\Slash{k_2}\Slash{\ell_2}\Slash{k_4}\Slash{\ell_1}))^4
\over (2\ell_1\cdot k_1)^2 (2\ell_1\cdot k_2)
(2 \ell_1\cdot k_3) (2\ell_1\cdot k_4)}\cr}
\anoneqn
$$
Where $\spb{a}.b\spa{b}.c\spb{c}.d\cdots \spa{m}.a
=\tr_{-} ( \Slash{a}\Slash{b}\Slash{c}\cdots\Slash{m} )$
and $\tr_{\pm}( \rho) =\tr ( {(1\pm\gamma_5)\over 2})  \rho)$
(We must also include the contribution with the other choice of helicities on
the internal legs:
$$
A(\phi_1 , \phi_2 , \ell^+_1 , \ell^-_2 )\, A(\phi_3 , \phi_4 , \ell^+_2 ,
\ell^-_1)
\anoneqn
$$
This is equivalent to setting $\ell_2 \rightarrow -\ell_1$ and
$\ell_1 \rightarrow
-\ell_2$ )

If we combine these two contributions and carry out the integrations
(which a rather more complicated than those for the scalar case)
we find
$$
\eqalign{{\kappa^4\rg\over(4\pi)^{2-\eps}}
\biggl(
&{\left (u^{2}-ut+t^{2}\right )\over8}\ln (-s)^{2}-{ {{1\over240}\,
\left (u^{2}+41\,ut+t^{2}\right )\ln
(-s)}}
\cr &\qquad+{t^{3}\ln (-t)\ln (-s)\over 4\,s}+{u^{3}\ln (-u)\ln
(-s)\over 4\,s}\biggr) \cr }
\anoneqn
$$
If we add this to the  result for intermediate scalars
we obtain
$$
\eqalign{{\kappa^4\rg\over(4\pi)^{2-\eps}} \biggl(
&{ {\left (u^{4}+2\,u^{3}t+2\,ut^{3}+3\,u
^{2}t^{2}+t^{4}\right )\ln (-s)}\over{2\,ut}}{1\over\epsilon}\cr
&\qquad+{1\over4}\left (3
\,u^{2}+2\,ut+3\,t^{2}\right ){\rg}\,\ln (-s)^{2}\cr
&\qquad\qquad+{
(  t^4 + s^4 )
\ln (-s)\ln (-t)\over{2\,ts}}
+{{\left (
 s^4 +  u^4
\right )
\ln (-s)\ln (-u)}\over{2\,us}}
\cr
&\qquad\qquad\qquad\qquad+{1\over240}{
{\left
(-163\,u^{3}t-43\,u^{2}t^{2}-163\,ut^{3}\right )
\ln (-s)}\over{ut}}
 \biggr) \cr}
\anoneqn
$$
{}From this we can write down an expression with the correct
cuts in all channels
$$
\eqalign{
8(s^4+t^4)& I_4(s,t)+8(s^4+u^4)I_4(s,u)+8(u^4+t^4)I_4(t,u)
\cr
&
-8s(3s^2+t^2+u^2) I_3(s)
-8t(3t^2+s^2+u^2) I_3(t)
-8u(3u^2+t^2+s^2) I_3(u)
\cr
&
+{2 (163u^2+163t^2 +43tu) \over 15 } I_2(s)
+{2 (163u^2+163s^2 +43us) \over 15 } I_2(t)
\cr
&+{2 (163s^2+163t^2 +43ts) \over 15 } I_2(u)
\cr}
\anoneqn
$$
This expression has the correct cuts in all channels and contains the correct
infinite pieces. It is {\it not} the correct full result.  As discussed in the
previous section we expect the full answer to differ from the above by finite,
non-logarithmic rational polynomials in the momentum invariants. However the
above expression does contain the correct $1/\eps$ divergences.

Contained in the above we expect a IR infinity of the form
$$
\eqalign{
& {i\rg \kappa^2 \over2(4\pi)^{2-\epsilon}}
{ \big((-s)^{1-\epsilon}+(-t)^{1-\epsilon}+(-u)^{1-\epsilon}\big)
\over \epsilon^2}
A^{\rm tree}(\phi_1,\phi_2,\phi_3,\phi_4)
\cr
&=
{\rg \kappa^4 \over2(4\pi)^{2-\epsilon}}
{ \big((-s)^{1-\epsilon}+(-t)^{1-\epsilon}+(-u)^{1-\epsilon}\big)
\over \epsilon^2}
\Bigl(
{ s^2 +t^2 \over 2u}
+{ s^2+u^2 \over 2t}
+{ t^2+u^2 \over 2s}
\Bigr)
\cr}
\anoneqn
$$
By examination we can identify this term plus an additional
infinity,
$$
-{203\over 160\eps }{\left(t^{2}+s^2+u^{2}\right)}
\anoneqn
$$
Since the IR divergences are accounted for, this must be a UV infinity.
By inspection, we can see this corresponds to the counter term in the
Lagrangian in eqn.~(\use\ScalarCounter) - with the correct coefficient.

Before finishing this section we can consider the case where gravity is
couple to a set of $N$ scalar $\phi_i$. Using the techniques above we
can identify the counter Lagrangian to be
$$
\eqalign{
\Delta{\cal L}
= &{\sqrt{g}\over\epsilon}
\bigg\{\sum_{i=1}^N
\left( { 202 +N \over 80 } \right) ( \partial_\mu\phi_i\partial^\mu\phi_i)^2 )
\cr &
\sum_{i \neq j } \bigg(
\left({N-198 \over960}\right)(
\partial_\mu\phi_i\partial^\mu\phi_i)
(\partial_\nu\phi_j\partial^\nu\phi_j)
\cr&
\qquad\qquad\qquad\qquad\qquad\qquad+
\sum_{i \neq j }
\left({N+402 \over480}\right)
(\partial_\mu\phi_i\partial_\nu\phi_i)
(\partial^\mu\phi_j\partial^\nu\phi_j)
\bigg)\bigg\}
\cr
}
\anoneqn
$$
It is possible to examine the counter terms for a variety of
types of matter coupled to gravity. Further examples will be given
elsewhere [\use\Paul].

\section{The Cut Calculation of
$\Aloop(1^+,2^+,3^+,4^+)$ with a massive scalar.}

Although a naive interpretation the Cutkosky rules suggests that
one may only use them to evaluate the amplitudes up to finite
rational polynomials in the momentum invariants it may be possible to
use the rules to evaluate these also. The idea is fairly simple.
In dimensional regularisation, if we can evaluate the cuts to
order $\eps^n$ then we can reconstruct the rational
polynomial terms to
order $\eps^{n-1}$. The difficulty lies in
evaluating the cuts consistently to all orders. That this is possible,
has been demonstrated by Bern and Morgan in ref.~[\use\BernMorgan]
where it was shown how to, with care, evaluate the cuts to higher order
in
$\eps$.
Previously we have used the on-shell tree amplitudes with intermediate
legs having momenta
in $D=4$. However the loop momentum integral has momenta in
$D=4-2\eps$. This involves an error, which although not contributing
to finite order in $\ln(-s)$ [\use\SusyOne], gives an error in terms
$\eps \ln(-s)$ which feeds down to finite polynomial terms.
To correctly carry out the $D=4-2\eps$ loop momentum integrals it is
convenient to split the $D$-dimensional momentum into a
four dimensional and a $-2\eps$ momentum, $\mu$,  whence
$$
p^2_{D} = p_4^2 -\mu^2
\anoneqn
$$
and the integration splits up as
$$
d^Dp \rightarrow  d^4 p \; d^{-2\eps} \mu
\eqn\RegulateInt
$$
This prescription is well described in ref.~[\use\Mahlon].
If calculating the amplitude for a massless scalar circulating in the loop
the prescription is now clear: one uses on-shell four dimensional
amplitudes but where the scalar now has a mass $\mu^2$ and
one integrates according to eqn.~(\use\RegulateInt).
Hopefully an example will make this clear.
The first example we consider will be the contribution to
four graviton scattering when the external gravitons all have the same
helicity from a massless scalar  circulating in the loop,
$$
\Aloop_{\rm scalar} (1^+,2^+,3^+,4^+)
\anoneqn
$$
\footnote{${}^{\dagger}$}{Due to supersymmetric Ward identities [\use\Susy]
this amplitude due to massless scalars circulating is equal to that for
gravitons circulating.}
This can be found in ref.~[\use\GravityString] where it was calculated
using string-based diagrammatic rules. The answer is a finite
rational polynomial and hence we might expect
it not to be calculable using cuts.
This is consistent with the fact that the tree amplitudes necessary to
evaluate the cuts $\Atree(1^+,2^+,\phi,\phi)$ vanish for a massless
scalar. However if we consider the tree amplitude for a scalar with
A
mass $\mu^2$ we find,
$$
\eqalign{
A_4(1^s,2^+,3^+,4^s)
&= -i {\kappa^2  \over 4} {(\mu^4) \spb2.3^2 \over\spa2.3^2}
\Bigl( {1 \over (k_1+k_2)^2 -\mu^2 }+{1 \over (k_1+k_3)^2-\mu^2 }\Bigr)\cr
}
\anoneqn
$$
When we calculate the cut in the, for example,
$s_{12}$ channel  the two factors yield two box integrals
with ordering of legs $1234$ and $1243$. The $1234$ ordering will contribute
(after inserting the factors of $1/(\ell_1^2-\mu^2)$ and
$1/(\ell_2^2-\mu^2)$)
$$
\eqalign{
{\kappa^4  \over 16 } { \spb1.2^2\spb3.4^2 \over \spa1.2^2\spa3.4^2}&
\int {d^4p \over (2 \pi)^{4 -2\eps}}
 d^{-2 \eps}\mu
\cr
\times & \; {\mu^8 \over
(p^2 -\mu^2) ((p-k_1)^2 - \mu^2)
((p - k_1 - k_2)^2 - \mu^2)
((p - k_1 - k_2 - k_3)^2 - \mu^2) } \; .
\cr}
\eqn\MasslessCase
$$
The integrals over $-2\eps$ loop momenta are evaluated using
eqn.(22) of ref.~[\use\Mahlon].
$$
A
\int { d^{-2\eps} \mu  \over (2\pi)^{-2\eps} }
=
A
{-\eps (4\pi)^{\eps} \over \Gamma(1-\eps)}
\int_{0}^{\infty}
d\mu^2 (\mu^2)^{-1-\eps}
\anoneqn
$$
and (after Feynman parameterising the $\int d^4 p$ integral )
$$
 \int_0^\infty d\mu^2 (\mu^2)^{-1 - \eps}
{ \mu^8 \over
( s a_1 a_3 + t a_2 a_4 - \mu^2)^2 }
=
{\pi (3-\eps ) \over
\sin (\pi \eps )}
(- s a_1 a_3 - t a_2 a_4 )^{2-\eps}
\anoneqn
$$
We can then reconstruct the cut in the $s$-channel
$$
- {i\kappa^4  \over 16 (4\pi)^{2-\eps} }
{ \spb1.2^2\spb3.4^2 \over \spa1.2^2\spa3.4^2}
{\pi \eps (3-\eps ) \over
\sin (\pi \eps )}
{ \eps (1-\eps)(2-\eps) \over  \Gamma(1-\eps) \Gamma(1+\eps)}
\biggl( I^{D=12-2\eps}_{1234}
\; + I^{D=12-2\eps}_{1243}
\biggr)
\anoneqn
$$
where the definition of the $D$-dimensional box integral is
$$
I^D
\equiv
\Gamma(4-D/2) \int d a_i\;
\delta({\textstyle \sum} a_i - 1)
{ 1  \over
(- s a_1 a_3 - t a_2 a_4 )^{4-D/2}}
\anoneqn
$$
By symmetry we can deduce that the following object has the correct all-
order in $\eps$ cuts in all three channels and hence must be the
correct  amplitude,
$$
- {i\kappa^4  \over 16}
{ \spb1.2^2\spb3.4^2 \over \spa1.2^2\spa3.4^2}
{\eps \over(4\pi)^{2-\eps} \Gamma(1-\eps)}
{\pi (3-\eps ) \over
\sin (\pi \eps )}
{ \eps (1-\eps)(2-\eps) \over   \Gamma(1+\eps)}
\biggl( I^{D=12-2\eps}_{1234}
\; + I^{D=12-2\eps}_{1243}
\; + I^{D=12-2\eps}_{1323}\biggr)
\anoneqn
$$
Since this is now the correct answer to all order in $\eps$
A
we can evaluate it merely at order $\eps^0$.
Thus we need $I^{D=12-2\eps}$ to order
$\eps^{-1}$
$$
\eqalign{
I^{D=12-2\eps}
&=\Gamma(-2+\eps) \int d a_i\;
\delta({\textstyle \sum} a_i - 1)
(- s a_1 a_3 - t a_2 a_4 )^{2-\eps}
\cr
&=
{1 \over 2\eps } \int d a_i\;
\delta({\textstyle \sum} a_i - 1)
(- s a_1 a_3 - t a_2 a_4 )^{2}
+O(\eps^0)
\cr
&= {  2s^2 +st +2t^2 \over 5040 \eps } +O(\eps^0)
\cr}
\anoneqn
$$
and thus the amplitude is
$$
\eqalign{
&{i\kappa^4  \spb1.2^2\spb3.4^2 \over \spa1.2^2\spa3.4^2}
{ 6 \over 16(4\pi)^2 }
 { 4s^2 +4t^2+4u^2 +st+tu+su \over 2520 }
\cr
& \hskip 1.0 truecm
=
{i\kappa^4  \over (4\pi)^2 }
\Bigl( { st \over \spa1.2\spa2.3\spa3.4\spa4.1} \Bigr)^2
{(s^2 +t^2+u^2) \over  3840 }
\cr}
\anoneqn
$$
This reproduces the result of ref.~[\use\GravityString]
for a real scalar. (For a complex scalar multiply by a factor of 2).
and is consistent with [\use\Zak].
thus we have thus demonstrated, as was done for QCD in [\use\BernMorgan]
how the cuts can reproduce the finite
rational polynomials in amplitudes.

After demonstrating the validity of the method let us calculate
the amplitude with a massive scalar of mass $M$ in the loop.
This has been calculated in the $M\rightarrow \infty$ limit
previously [\use\LD]. This will involve replacing $\mu^2$ by $\mu^2+M^2$ in the
initial cut expression. Following the analysis to
eqn.~(\use\MasslessCase) $\mu^8$ is replaced by $(\mu^2+M^2)^4$ in the
numerator and $\mu^2$ by $\mu^2+M^2$ in the numerators. The various
terms can then be evaluated using
$$
\eqalign{
 \int d a_i\;
 \delta({\textstyle \sum}&  a_i - 1)
 \int_0^\infty d\mu^2 (\mu^2)^{-1 - \eps}
{ \mu^ {2n}\over
( s a_1 a_3 + t a_2 a_4 - \mu^2-M^2)^2 }
\cr
&=
{(-1)^{n} \pi (n-\eps-1 ) \over
\sin (\pi \eps )}
( s a_1 a_3 + t a_2 a_4-M^2 )^{n-\eps-2}
\cr
&=
{(-1)^{n} \pi (n-\eps-1 ) \over
\sin (\pi \eps )}
A
{1 \over \Gamma(2+\eps-n) }
I^{D=2n-2\eps+4}
\cr}
\anoneqn
$$
Putting the pieces together we may
obtain
$$
\eqalign{
- {i\kappa^4  \over 16 (4\pi)^{2-\eps} }
{ \spb1.2^2\spb3.4^2 \over \spa1.2^2\spa3.4^2}
& {\pi\eps  \over
\sin (\pi \eps )\Gamma(1-\eps) \Gamma(1+\eps) }
\biggl(
{ \eps (1-\eps)(2-\eps)(3-\eps )}
I^{D=12-2\eps}_{1234}
\cr
&\;\;+4M^2 \eps (1-\eps)(2-\eps)
I^{D=10-2\eps}_{1234}
+6M^4  \eps (1-\eps)I^{D=8-2\eps}_{1234}
\cr
&\;\;+4M^6 \eps I^{D=6-2\eps}_{1234}
-M^8  I^{D=4-2\eps}_{1234}
+{\rm \{1243\} }
+{\rm \{1324\} }
\biggl)
\cr}
\anoneqn
$$
which will be valid to all orders in $\eps$.
To evaluate to $\eps^0$,
we will only need the infinite parts of the
$D=12$,$D=10$ $D=8$ and $D=6$ integrals and the
finite part of the $D=4$ integral.
(The $D=6$ integral will drop out to order $\eps^0$ since
the box is both UV and IR finite in $D=6$.)
We can evaluate these objects to $O(\eps^0)$, using
$$
\eqalign{
{ \eps (1-\eps)(2-\eps) \over \Gamma(1+\eps) }
I^{D=12-2\eps}_{1234}
A
&=
\int d a_i\;
\delta({\textstyle \sum} a_i - 1)
( s a_1 a_3 + t a_2 a_4-M^2 )^{2-\eps}
\cr
&=
{ 2s^2 +st +2t^2 \over 2520 }
-{ M^2 (s+t) \over 60 }
+{ M^4 \over 6 } \cr
{ \eps (1-\eps) \over \Gamma(1+\eps) }
I^{D=10-\eps}_{1234}
&=\int d a_i\;
\delta({\textstyle \sum} a_i - 1)
( s a_1 a_3 + t a_2 a_4-M^2)^{1-\eps}
={ s+t \over 120 }  -{ M^2 \over 6 }
\cr
{ \eps  \over \Gamma(1+\eps) }
I^{D=8-2\eps}_{1234}
&=\int d a_i\;
\delta({\textstyle \sum} a_i - 1)
( s a_1 a_3 + t a_2 a_4-M^2 )^{-\eps}=
1/6
 \cr}
\anoneqn
$$
which will yield for the amplitude (for a real scalar),
$$
\eqalign{
- {i\kappa^4  \over 16 (4\pi)^2 }
\biggl( { st \over \spa1.2\spa2.3\spa3.4\spa4.1} \biggr)^2
\biggl(
&
{ (s^2 +t^2 +u^2 ) \over 240}
\;   +{1\over 2} M^4
\cr
&
\hskip 0.5truecm -M^8  (
I^{D=4}_{1234}
\;
A
+I^{D=4}_{1243}
\;
+I^{D=4}_{1324}
)
\biggl)
\cr}
\anoneqn
$$
This amplitude contains logarithms and dilogarithms from $I_{1234}^{D=4}$.
These terms could be correctly obtained from the cuts naively but we also have
the additional polynomial terms to give the full correct result. We might also
expect that the amplitude should contain terms of the form $\ln(m^2)$, which
would not have been recovered in the above analysis.  However, we can deduce
that no such terms appear by applying the arguments outlined in
ref.~[\BernMorgan]:  Such terms only appear with specific divergent
contributions; since we know that the amplitude is UV and IR finite we know
that no $\ln(m^2)$ terms will be found. The large $M$ expansion, on
$I^{D=4}_{1234}$ is
$$ \eqalign{ \int d a_i\; &\delta({\textstyle \sum} a_i - 1)
( s a_1 a_3 + t a_2 a_4-M^2 )^{-2-\eps}
\cr
&=\int d a_i\;
\delta({\textstyle \sum} a_i - 1)
(M^2)^{-2-\eps}
\biggl(
1
-(2+\eps){ S\over M^2 }
+(2+\eps)(3+\eps){ S^2 \over 2 M^4}
\cr
&-( 2+\eps)(3+\eps)(4+\eps){ S^3 \over 6 M^6 }
+( 2+\eps)(3+\eps)(4+\eps)(5+\eps) { S^4 \over 24 M^8 }
\cr
&-( 2 +\eps)(3+\eps)(4+\eps)(5+\eps)(6+\eps) { S^5 \over 120 M^{10} }
 \cdots \; \biggr)
\cr
&=
{1 \over 6 M^4 }
-{ s+t \over 60 M^6 }
+{ 2s^2 +st +2t^2 \over 840 M^8 }
-{ 3s^3 +st^2+s^2t +3 t^3 \over 7560 M^{10} }
\cdots \cr}
\anoneqn
$$
where $S=sa_1a_3+ta_2a_4$. From this we obtain
the large mass expansion for the amplitude
$$
\eqalign{
- {i\kappa^4  \over 16 (4\pi)^2 }
\biggl( { st \over \spa1.2\spa2.3\spa3.4\spa4.1} \biggr)^2
\biggl(
{ s t u \over 504  M^2 }
-{ (s^2+t^2+u^2)^2  \over  15120 M^4 }
&
-{ st u (s^2+t^2+u^2 ) \over 15840 M^6 }
\cdots
\biggr)
\cr}
\anoneqn
$$
This amplitude tends to zero as $M\rightarrow \infty$ as expected with
the $1/M^2$  term having the same coefficient as calculated previously
[\use\LD].

\section{Conclusions}

In perturbative gravity and gauge theories the calculation of
amplitudes is a painful process where huge intermediate expression
eventually collapse to relatively small final answers.  This
computational complexity will remain a challenge despite considerable
improvement in both techniques and algebraic computing facilities.
Any technique which attempts to avoid this
computational explosion must be explored carefully.
Unitarity in the form of the
Cutkosky rules is one such technique. By sewing together tree amplitudes upon
which much simplifications have already been performed one can build upon
previous calculations and avoid part of the algebraic complexity.
Unfortunately, the full amplitude is not calculated by the Cutkosky rules
but
the rules are a powerful constraint upon the form of amplitudes.
In some cases unitarity is enough to fix the
the amplitudes completely, however naively applying the Cutkosky rules
in the general case leaves ambiguities.  These ambiguities however can
be expressed in such a form that they only effect the finite terms
polynomial in momentum invariants. Thus, in principle, one can
calculate the infinities using the Cutkosky rules.
More sophisticated use of the Cutkosky rules [\use\BernMorgan]
can also be used to determine the entire structure of a loop amplitude.
In quantum gravity it is most often the infinities in amplitudes which
are of more interest rather than the finite terms (The case in QCD is often
the other way around.).
Here we have shown
in several cases how the Cutkosky rules can be practically used to evaluate
infinities. In any massless theory there are both IR and UV
infinities. Although the cuts do not distinguish between these
sources, the general structure of the IR infinities can be determined
[\use\LD].
With this information we can isolate the UV divergences.  We have
explicitly shown how the UV infinities in scalar coupled matter appear
in an amplitude using this technique and are in agreement with the
known results which were obtained by a very different route. Other
infinities previously unknown may be obtained in this manner
[\use\Paul].

We would like to thank Zvi Bern and Andrew Morgan for useful discussions
regarding the use of Cutkosky rules for massive particles,
Christian Schubert for discussions relating to the world-line formalism
for gravity and Lance Dixon
for discusions both regarding the possibility of disentangling the UV and IR
infinities
within the amplitudes
and for
access to his calculations regarding massive scalar amplitudes.
This research was supported by PPARC.

\listrefs

\end